

%
\newbox\leftpage \newdimen\fullhsize \newdimen\hstitle \newdimen\hsbody
\tolerance=1000\hfuzz=2pt
\def\bigans{b }
\message{ big or little (b/l)? }\read-1 to\answ
\ifx\answ\bigans\message{(This will come out unreduced.}
\magnification=1200\baselineskip=18pt plus 2pt minus 1pt
\hsbody=\hsize \hstitle=\hsize 
\else\def\apans{l }\message{ lyman or hepl (l/h) (lowercase!) ? }
\read-1 to \apansw\message{(This will be reduced.}
\let\lr=L
\magnification=1000\baselineskip=16pt plus 2pt minus 1pt
\voffset=-.31truein\vsize=7truein
\hstitle=8truein\hsbody=4.75truein\fullhsize=10truein\hsize=\hsbody
\ifx\apansw\apans\special{ps: landscape}\hoffset=-.59truein
  \else\hoffset=.05truein\fi
\output={\ifnum\pageno=0 
  \shipout\vbox{\hbox to \fullhsize{\hfill\pagebody\hfill}}\advancepageno
  \else
  \almostshipout{\leftline{\vbox{\pagebody\makefootline}}}\advancepageno
  \fi}
\def\almostshipout#1{\if L\lr \count1=1
      \global\setbox\leftpage=#1 \global\let\lr=R
  \else \count1=2
    \shipout\vbox{\ifx\apansw\apans\special{ps: landscape}\fi 
      \hbox to\fullhsize{\box\leftpage\hfil#1}}  \global\let\lr=L\fi}
\fi
%
\catcode`\@=11 
\newcount\yearltd\yearltd=\year\advance\yearltd by -1900

\def\Title#1#2{\nopagenumbers\abstractfont\hsize=\hstitle\rightline{#1}%
\vskip 1in\centerline{\titlefont #2}\abstractfont\vskip .5in\pageno=0}
\def\Date#1{\vfill\leftline{#1}\tenpoint\supereject\global\hsize=\hsbody%
\footline={\hss\tenrm\folio\hss}}
\def\draftmode{\def\draftdate{{\rm preliminary draft:
\number\month/\number\day/\number\yearltd\ \ \hourmin}}%
\headline={\hfil\draftdate}\writelabels\baselineskip=20pt plus 2pt minus 2pt
{\count255=\time\divide\count255 by 60 \xdef\hourmin{\number\count255}
	\multiply\count255 by-60\advance\count255 by\time
   \xdef\hourmin{\hourmin:\ifnum\count255<10 0\fi\the\count255}}}

\def\nolabels{\def\eqnlabel##1{}\def\eqlabel##1{}\def\reflabel##1{}}
\def\writelabels{\def\eqnlabel##1{%
{\escapechar=` \hfill\rlap{\hskip.09in\string##1}}}%
\def\eqlabel##1{{\escapechar=` \rlap{\hskip.09in\string##1}}}%
\def\reflabel##1{\noexpand\llap{\string\string\string##1\hskip.31in}}}
\nolabels
%
\global\newcount\secno \global\secno=0
\global\newcount\meqno \global\meqno=1
\def\newsec#1{\global\advance\secno by1
\xdef\secsym{\the\secno.}\global\meqno=1
\bigbreak\bigskip
\noindent{\bf\the\secno. #1}\par\nobreak\medskip\nobreak}
\xdef\secsym{}
\def\appendix#1#2{\global\meqno=1\xdef\secsym{\hbox{#1.}}\bigbreak\bigskip
\noindent{\bf Appendix #1. #2}\par\nobreak\medskip\nobreak}
%
%
\def\eqnn#1{\xdef #1{(\secsym\the\meqno)}%
\global\advance\meqno by1\eqnlabel#1}
\def\eqna#1{\xdef #1##1{\hbox{$(\secsym\the\meqno##1)$}}%
\global\advance\meqno by1\eqnlabel{#1$\{\}$}}
\def\eqn#1#2{\xdef #1{(\secsym\the\meqno)}\global\advance\meqno by1%
$$#2\eqno#1\eqlabel#1$$}
%
\newskip\footskip\footskip14pt plus 1pt minus 1pt 
\def\f@@t{\baselineskip\footskip\bgroup\aftergroup\@foot\let\next}
\setbox\strutbox=\hbox{\vrule height9.5pt depth4.5pt width0pt}
\global\newcount\ftno \global\ftno=0
\def\foot{\global\advance\ftno by1\footnote{$^{\the\ftno}$}}
%
%
\global\newcount\refno \global\refno=1
\newwrite\rfile
\def\ref{[\the\refno]\nref}
\def\nref#1{\xdef#1{[\the\refno]}\ifnum\refno=1\immediate
\openout\rfile=refs.tmp\fi\global\advance\refno by1\chardef\wfile=\rfile
\immediate\write\rfile{\noexpand\item{#1\ }\reflabel{#1}\pctsign}\findarg}
\def\findarg#1#{\begingroup\obeylines\newlinechar=`\^^M\pass@rg}
{\obeylines\gdef\pass@rg#1{\writ@line\relax #1^^M\hbox{}^^M}%
\gdef\writ@line#1^^M{\expandafter\toks0\expandafter{\striprel@x #1}%
\edef\next{\the\toks0}\ifx\next\em@rk\let\next=\endgroup\else\ifx\next\empty%
\else\immediate\write\wfile{\the\toks0}\fi\let\next=\writ@line\fi\next\relax}}
\def\striprel@x#1{} \def\em@rk{\hbox{}} {\catcode`\%=12\xdef\pctsign{
\def\semi{;\hfil\break}
\def\addref#1{\immediate\write\rfile{\noexpand\item{}#1}} 
\def\listrefs{\vfill\eject\immediate\closeout\rfile
\baselineskip=12pt\centerline{{\bf References}}\bigskip{\frenchspacing%
\escapechar=` \input refs.tmp\vfill\eject}\nonfrenchspacing}
\def\startrefs#1{\immediate\openout\rfile=refs.tmp\refno=#1}
\def\figures{\centerline{{\bf Figure Captions}}\medskip\parindent=40pt}
\def\fig#1#2{\medskip\item{Figure ~#1:  }#2}
\catcode`\@=12 
%
\ifx\answ\bigans
\font\titlerm=cmr10 scaled\magstep3 \font\titlerms=cmr7 scaled\magstep3
\font\titlermss=cmr5 scaled\magstep3 \font\titlei=cmmi10 scaled\magstep3
\font\titleis=cmmi7 scaled\magstep3 \font\titleiss=cmmi5 scaled\magstep3
\font\titlesy=cmsy10 scaled\magstep3 \font\titlesys=cmsy7 scaled\magstep3
\font\titlesyss=cmsy5 scaled\magstep3 \font\titleit=cmti10 scaled\magstep3
\else
\font\titlerm=cmr10 scaled\magstep4 \font\titlerms=cmr7 scaled\magstep4
\font\titlermss=cmr5 scaled\magstep4 \font\titlei=cmmi10 scaled\magstep4
\font\titleis=cmmi7 scaled\magstep4 \font\titleiss=cmmi5 scaled\magstep4
\font\titlesy=cmsy10 scaled\magstep4 \font\titlesys=cmsy7 scaled\magstep4
\font\titlesyss=cmsy5 scaled\magstep4 \font\titleit=cmti10 scaled\magstep4
\font\absrm=cmr10 scaled\magstep1 \font\absrms=cmr7 scaled\magstep1
\font\absrmss=cmr5 scaled\magstep1 \font\absi=cmmi10 scaled\magstep1
\font\absis=cmmi7 scaled\magstep1 \font\absiss=cmmi5 scaled\magstep1
\font\abssy=cmsy10 scaled\magstep1 \font\abssys=cmsy7 scaled\magstep1
\font\abssyss=cmsy5 scaled\magstep1 \font\absbf=cmbx10 scaled\magstep1
\skewchar\absi='177 \skewchar\absis='177 \skewchar\absiss='177
\skewchar\abssy='60 \skewchar\abssys='60 \skewchar\abssyss='60
\fi
\skewchar\titlei='177 \skewchar\titleis='177 \skewchar\titleiss='177
\skewchar\titlesy='60 \skewchar\titlesys='60 \skewchar\titlesyss='60
\def\titlefont{\def\rm{\fam0\titlerm}
\textfont0=\titlerm \scriptfont0=\titlerms \scriptscriptfont0=\titlermss
\textfont1=\titlei \scriptfont1=\titleis \scriptscriptfont1=\titleiss
\textfont2=\titlesy \scriptfont2=\titlesys \scriptscriptfont2=\titlesyss
\textfont\itfam=\titleit \def\it{\fam\itfam\titleit} \rm}
\ifx\answ\bigans\def\abstractfont{\tenpoint}\else
\def\abstractfont{\def\rm{\fam0\absrm}
\textfont0=\absrm \scriptfont0=\absrms \scriptscriptfont0=\absrmss
\textfont1=\absi \scriptfont1=\absis \scriptscriptfont1=\absiss
\textfont2=\abssy \scriptfont2=\abssys \scriptscriptfont2=\abssyss
\textfont\itfam=\bigit \def\it{\fam\itfam\bigit}
\textfont\bffam=\absbf \def\bf{\fam\bffam\absbf} \rm} \fi
\def\tenpoint{\def\rm{\fam0\tenrm}
\textfont0=\tenrm \scriptfont0=\sevenrm \scriptscriptfont0=\fiverm
\textfont1=\teni  \scriptfont1=\seveni  \scriptscriptfont1=\fivei
\textfont2=\tensy \scriptfont2=\sevensy \scriptscriptfont2=\fivesy
\textfont\itfam=\tenit \def\it{\fam\itfam\tenit}
\textfont\bffam=\tenbf \def\bf{\fam\bffam\tenbf} \rm}
%
%
\def\noblackbox{\overfullrule=0pt}
\hyphenation{anom-aly anom-alies coun-ter-term coun-ter-terms}
\def\inv{^{\raise.15ex\hbox{${\scriptscriptstyle -}$}\kern-.05em 1}}
\def\dup{^{\vphantom{1}}}
\def\Dsl{\,\raise.15ex\hbox{/}\mkern-13.5mu D} 
\def\dsl{\raise.15ex\hbox{/}\kern-.57em\partial}
\def\del{\partial}
\def\Psl{\dsl}
\def\tr{{\rm tr}} \def\Tr{{\rm Tr}}
\font\bigit=cmti10 scaled \magstep1
\def\biglie{\hbox{\bigit\$}} 
\def\lspace{\ifx\answ\bigans{}\else\qquad\fi}
\def\lbspace{\ifx\answ\bigans{}\else\hskip-.2in\fi} 
\def\boxeqn#1{\vcenter{\vbox{\hrule\hbox{\vrule\kern3pt\vbox{\kern3pt
	\hbox{${\displaystyle #1}$}\kern3pt}\kern3pt\vrule}\hrule}}}
\def\mbox#1#2{\vcenter{\hrule \hbox{\vrule height#2in
		\kern#1in \vrule} \hrule}}  
%
\def\CAG{{\cal A/\cal G}}   
\def\CA{{\cal A}} \def\CC{{\cal C}} \def\CF{{\cal F}} \def\CG{{\cal G}}
\def\CL{{\cal L}} \def\CH{{\cal H}} \def\CI{{\cal I}} \def\CU{{\cal U}}
\def\CB{{\cal B}} \def\CR{{\cal R}} \def\CD{{\cal D}} \def\CT{{\cal T}}
\def\e#1{{\rm e}^{^{\textstyle#1}}}
\def\grad#1{\,\nabla\!_{{#1}}\,}
\def\gradgrad#1#2{\,\nabla\!_{{#1}}\nabla\!_{{#2}}\,}
\def\ph{\varphi}
\def\psibar{\overline\psi}
\def\om#1#2{\omega^{#1}{}_{#2}}
\def\vev#1{\langle #1 \rangle}
\def\lform{\hbox{$\sqcup$}\llap{\hbox{$\sqcap$}}}
\def\darr#1{\raise1.5ex\hbox{$\leftrightarrow$}\mkern-16.5mu #1}
\def\lie{\hbox{\it\$}} 
\def\ha{{1\over2}}
\def\half{{\textstyle{1\over2}}} 
\def\roughly#1{\raise.3ex\hbox{$#1$\kern-.75em\lower1ex\hbox{$\sim$}}}

\font\names=cmbx10 scaled\magstep1

\baselineskip=20pt
\Title{PUP-TH-1408}
{\vbox{\centerline
{A Simple Test }
\centerline{
for Non-Gaussianity in CMBR Measurements}}}
\font\large=cmr10 scaled\magstep3
\font\names=cmbx10 scaled\magstep1
\centerline{\bf\names Paul Graham$$}
\centerline{\bf\names Neil Turok$$}
\centerline{Joseph Henry Laboratories,  Princeton University}
\centerline{Princeton, NJ 08544.}
\centerline{\bf\names P. M. Lubin$$}
\centerline{\bf\names J. A. Schuster$$}
\centerline{Physics Department, University of California at Santa Barbara}
\centerline{Santa Barbara, CA 93106.}

\bigskip

\centerline{\bf Abstract}
\baselineskip=12pt
\smallskip
We propose a set of statistics $S_q$ for detecting non-gaussianity in
CMBR anisotropy data sets.  These statistics
are both simple and, according to calculations over a space
of linear combinations of three-point functions, nearly optimal at
detecting certain types of non-gaussian features.  We apply $S_3$ to
the UCSB SP91 experiment and find that the mean of
the four frequency channels is by this criterion
strongly non-gaussian. Such an observation would be highly
unlikely in a gaussian theory with a small coherence
angle, such as standard
($n=1$,
$\Omega_b = .05$, $h = .5$, $\Lambda=0$)
inflation.
 We cannot conclude
that the non-gaussianity is cosmological in origin, but if we assume it
due instead to foreground contamination or instrumental effects, and remove
the points which are
clearly responsible for the non-gaussian behavior, the rms of the remaining
fluctuations is too small for consistency with standard
inflation at high confidence.
Further data are clearly
needed however, before definitive conclusions may be drawn.
We also generalize the ideas behind this statistic to
non-gaussian features that might be detected
 in
other experimental schemes.

\medskip
\noindent {{\it subject headings:} \quad cosmic microwave background radiation,
\quad cosmology}

\Date{June 1993}

\baselineskip=16pt
\bigskip
\centerline{\bf 1. Introduction}
\smallskip
Many experiments, current and proposed, are dedicated to measuring
fluctuations in the Cosmic Microwave Background Radiation (CMBR).  These
measurements promise a strong experimental test of theories  of structure
formation in the early universe, as each theory predicts a distinct
magnitude and form for CMBR fluctuations. In  inflationary models,
the structure generation mechanism is linear, resulting  in
a gaussian pattern of fluctuations, completely characterised by
its power spectrum (see e.g. Efstathiou 1990).
By contrast, in  theories based on symmetry breaking and field ordering
 (e.g. cosmic strings and textures) nonlinear dynamics lead to
a non-gaussian anisotropy pattern,
due in part to horizon-sized topological defects at the
epoch of last scattering
(Kaiser \& Stebbins 1984; Turok \& Spergel 1990; Bennett \& Rhie 1992;
Coulson, Pen, \& Turok 1993; Pen, Spergel, \& Turok 1993).

CMBR measurements have not yet discriminated among different structure
formation theories to the extent that one might have hoped.
This is in part because the measurements are
still far from perfect.  COBE
has a low signal to noise ratio and large angular smoothing scale
(Smoot {\it et. al.} 1992; Ganga {\it et. al.} 1993)
, while other experiments are more accurate but cover only a small region
of
the sky (e.g.
Gaier {\it et. al.} 1992; Schuster, {\it et. al} 1993;
Meinhold {\it et. al.} 1993;
Gundersen {\it et. al.} 1993;  Cheng {\it et. al.} 1993).
But it is also because most theories
include parameters ($n,$ $h$, $\Omega$, $\Omega_B$,
$\Lambda$, (Tensor/Scalar)...) which can be adjusted
to modify the power spectrum. Such adjustments do not  however
alter the  more fundamental
 gaussian or non-gaussian character of the theories,
which
may be a more powerful discriminator.
This paper is aimed at finding statistics which focus
on this basic question. Other recent papers which propose
statistical tests for non-gaussianity
are Luo \& Schramm (1993) and Moessner, Perivaropoulos
\& Brandenberger (1993).

We are attempting
to extract information from very small data sets, obtained
in very difficult experiments. This is of course quite hazardous:
it is unlikely that
the idealized assumptions we
shall make about the experimental errors
are correct.
Any effect we see may well be due to foreground sources or systematic
instrumental effects, rather than non-gaussian cosmology.
Nevertheless it is an interesting exercise to
see how much may be learned, in principle, from experiments
of the type currently being undertaken. And this
may also serve as a guide to what kind of experimental effort  would
be most
informative in the future.
At the very least, we can make rigorous a process which is often performed
by eye: the identification of data points which are inconsistent with
gaussian
theories and must be thrown out as contaminated if these theories are to be
believed.

The various non-gaussian field ordering theories
predict different characteristic forms
on the microwave sky - for example, linelike discontinuities for
strings, hot and cold spots for textures. However they
all predict regions of sharp gradient, separated by
a characteristic scale of order
 the horizon at last scattering,  from one to a few degrees depending
on the reionisation history of the universe
(Pen {\it et. al.} 1993; Coulson {\it et. al.} 1993).
In this paper we shall discuss statistics which are sensitive to
degree scale large-gradient
 regions, and use them to discriminate between gaussian and non-gaussian
theories.

We shall follow tradition and
 use as our canonical `straw man' gaussian theory the
`standard' inflationary  model,
with parameters $n=1$, $h=.5$, $\Omega=1$, $\Omega_B = .05$,
$\Lambda=0$, and negligible tensor mode contribution.
This theory has a coherence angle of order $15'$, substantially
smaller than the scales degree scale experiments probe.
The results we get are similar to those obtained assuming
uncorrelated gaussian noise at each point differenced
on the sky. So while we shall find rather strong
evidence against such theories from the UCSB SP91 data,
we expect that
the constraints would be weaker for gaussian
theories with a large coherence angle, such as inflationary
theories
where there is a large tensor mode contribution (Crittenden
{\it et. al. } 1993).

As a concrete example of a degree scale CMBR measurement,
 we shall analyse the UCSB SP91
experiment
(Gaier {\it et. al.} 1992; Schuster {\it et. al.} 1993).
This is a `single difference' experiment; to generate a
single data point, the beam moves in a sinusoidal pattern, with the
antenna
temperature integrated antisymmetrically.  The result approximates the
first spatial
derivative of the fluctuations.  A set of results consist of nine to
fifteen data points (temperature differences), on
an arc of constant declination
 on the sky with 2.1 degrees separation between points.
To correct for atmospheric and other drifts, a best-fit line is removed.
  The
Schuster {\it et. al.} (1993) data set currently has the lowest error
per pixel ($5\times 10^{-6}$ for the four channel average) reported
for any CMBR anisotropy measurement.
Other ground-based experiments share many of these features, although
some integrate their intensities in such a way as to approximate the
second or third derivative of the fluctuations rather than the first and
are called double- or triple- difference experiments.
Depending on how many spatial
derivatives an experiment takes,  a gradient region will leave certain
`signature' forms on the data, as shown in Figure 1.

It might seem that a sharp gradient would span too few points to make
these
characteristic patterns visible, especially for higher derivatives with
complicated signatures.
However, the derivatives are taken by sweeping the beam on the
sky or by combining data from adjacent instrument positions.  These
methods give rise to an instrument response
function broad enough that the characteristic
signatures will be visible even for an infinitely sharp discontinuity in
the CMBR.

A final caveat should be added regarding our  use
of `classical' confidence intervals, as opposed to Bayesian
measures of the relative probability of different theories.
These are notoriously difficult to interpret.
We have done so mainly out of expedience -
it is far easier to construct realisations of gaussian theories
than for nongaussian theories. When enough non-gaussian maps are
available, the following
procedure may be preferable.
If one is comparing
theory $A$ to theory $B$, the change in the relative odds
following a measurement of a
continuous observable $X$ is given by the `Bayes factor'
\eqn\eaneil{
{P(X|A)\over P(X|B)}
}
where
probability of observing $X$ in the interval $dX$
is $P(X|A) dX$ according to  theory $A$ and similarly
for theory $B$.
As we mention in the conclusions, preliminary
results indicate that according to this, more
simply interpretable test,
the nongaussian theories may be significantly
favored by the UCSB SP91 data.

\bigskip
\centerline{\bf 2. Choosing Statistics}
\smallskip

All current theories of the origin of structure produce
fluctuations in the form of a stationary random process.
Any such process may be completely
characterized by the set of all $n$-point correlation
functions. On a one-dimensional data set these may be
estimated as
$$
C_{0\,r_1\,r_2 \cdots r_{n-1}}\equiv{1 \over (N - r_{n-1})}
\allowbreak \sum _{i=1} ^{N - r_{n-1}}
(x_i)(x_{i+r_1})\cdots (x_{i+r_{n-1}})
$$
$$
\hbox {where, by convention, }
 0 \leq r_1 \leq r_2 \leq \cdots \leq r_{n-1} \break
$$

We shall assume that the data set of interest has
zero mean, as in UCSB SP91, where a best-fit line is substracted as
explained above.  We shall adopt the convention
of normalizing the data set to unit variance, in order to
concentrate on the {\it shape} and not
the amplitude of the signal.

The one-point function $C_0$ is identically zero,
and the two-point function $C_{00}$ is identically
one (because of our unit-variance convention), and so the first nontrivial
correlation is $C_{0i}$, the two point
correlation function at scale $i$.  This does contain
information about the spectrum of fluctuations
--- it is the fourier transform of the power
spectrum --- but it is no help in distinguishing
gaussian from non-gaussian data.

The three-point function $C_{0ij}$ is a much more promising test for
non-gaussianity.  For gaussian noise, $<C_{0ij}>$ = 0 for all $i$ and $j$
(although
for finite data sets there will be random fluctuations about the expected
value of zero.)  For non-gaussian skies we expect non-zero three-point
functions.  For example, consider the three-point function $C_{000}$.
A data set containing a positive `bump' has several
outlying high points, and thus positive
skewness.  A downward-pointing bump will lead to
negative skewness.  Either way, the high absolute value of the
skewness could be
used to distinguish a data set drawn from a gaussian model from one drawn
from a region containing a bump (which, from fig. 1, is a likely
signature of non-gaussianity in a single-difference experiment.)

As we shall show,
skewness is not a very powerful statistic for reliably detecting
non-gaussianity in a noisy experiment.  We can improve on its performance
in two ways.

First, if the CMBR is non-gaussian,
a `bump' marking a gradient region may span two or more adjacent
points.  Even if the region of steep gradient on the sky were infinitely
sharp, it would register in at least two data points because of the
instrument's response function.
 Skewness fails to take advantage of these correlations among
closely neighboring points (obviously, since skewness is invariant under
spatial scrambling
of the data points.)  We can remedy this
shortcoming by combining several adjacent points; for example, to look for
bumps of width on the order of $q$, define
\eqn\eeesq{\eqalign{
S_q \equiv {1 \over (N-q+1)} \sum_{i=1}^{N-q+1}
\left({x_i + x_{i+1} + \cdots + x_{i+q - 1} \over q} \right) ^3
}}
This statistic responds much more sharply to several adjacent high
points than to the same number of high points scattered randomly over the
data
set, so it better distinguishes actual non-gaussian bumps from noise.  The
absolute value of $S_q$ will be near zero if
no bump exists and strongly non-zero if there is a single bump.

Of course, $S_q$ is not equal to a simple three-point function, but except
for the treatment of points near the edges of a data set, it {\it is}
equivalent to a linear
combination of three-point functions.
We will expand on this point later, when we show that $S_3$ is nearly
optimal,
among all linear combinations
of a certain set of three-point functions, at detecting bumps of width
near three.

A second way to improve the performance of almost {\it any} statistic
which detects bumps in a data set is to apply it not to the entire data
set but to shorter subsets or `windows'  of length $L\leq N$.  The final
statistic $S_{q;L}$ is defined as the absolute value of  the most extreme
(positive or negative) $S_q$ found in any of the $N - L + 1$ possible
window positions.  Use of these `sliding windows' improves the statistic's
performance for several reasons.  Most importantly, it prevents a positive
gradient region in one part of the data from cancelling a negative
gradient in another region (by isotropy, both signs are equally likely to
occur.)  The procedure also reduces the effects of noise on the statistic's
probability distribution by concentrating on only a few points around each
gradient signature.

If the data set contains a bump with some number $p$ of adjacent `strong'
(highly positive or highly negative) points, we expect the best results
when $L \approx p + 2(q-1)$.  This allows the window to contain every
group
of $q$ adjacent points which includes at least one `strong' point, and
no groups of $q$ points with no `strong' point.  $S_q$  is most
sensitive
to bumps with about $q$ 'strong' points.  We generally choose $p = q-1$, so
\eqn\eeel{\eqalign{
L=3(q-1)
}}

We will show that both in monte carlo runs and
on actual experimental data, statistics perform much better on sliding
windows of about
this scale than on entire data sets.
For long data sets, one might better consider the probability
distribution of $S_q$ over window positions, rather than the maximal
value, to
reduce sensitivity to a few extreme points.

Our favored choice of statistic will be $S_{3;6}$, as $q=3$ will be
sensitive to
bumps only slightly wider than the experimental response function, and
can thus detects gradient regions whose width (relative to the two-degree
scale set by the experiment) is fairly low but non-zero.  Equation \eeel\
then
sets the window length of $L=6$.

\bigskip
\centerline{\bf 3. Monte Carlo Results}
\smallskip

To test the relative power of different statistics, we devised a
monte carlo technique based on the UCSB SP91 experiment.
A large number of trial data sets $\{x_i:1 \leq i \leq N\}$ were
generated.
Typically, $N=13$ to match the UCSB SP91 experiment.  Half these data sets were
generated from a `null' gaussian model and half from a `bumpy'
non-gaussian model.

The null data sets could be generated in either of two ways. G.
Efstathiou (private communication 1993)
generously provided 1000 sets of 13 points, based on his
computer-generated `standard
 inflation-plus-CDM' skies and his simulation of the
properties of the UCSB SP91 experiment.  Alternatively, null sets could be
generated by
simply drawing $N$ independent, random points from a gaussian
distribution.  These
methods returned similar results (in that most interesting statistics
$S(\{x_i\})$ calculated from the $N$ points had similar distributions for
the two
null models.)  The statistics are apparently not greatly affected either
by correlations arising from the CMBR's power spectrum (no great
surprise, since the two-degree scale of this experiment is on the
low-frequency side of the power spectrum for this theory, and taking a
spatial derivative
further shifts the spectrum toward high spatial frequencies) or by
correlations arising from the instrument's response function (again no
surprise; the signature that we're looking for is symmetric and thus
orthogonal
to the antisymmetric response function.  Were we searching for point
sources rather that discontinuities, response--function--induced
correlations would be a more powerful confounding factor.)

The other half of the data sets were generated from a non-gaussian
`bumpy' model.  A single bump, centered at some random location $n_0$
within
the data set, was laid down:
$$
x_n = e^{-\alpha (n - n_0)^2}
$$
where we typically used $\alpha = 0.5$, corresponding to a bump with
full-width-half-max of 2.8 pixels.  Incidentally, $n_0$ is not necessarily
an integer; the center of the bump can lie between pixels. Independent
gaussian
noise was then added to each point to simulate instrument noise.
In both the null and `bumpy' models, the each data set $\{x_i\}$ was
normalized to zero mean and unit variance.

For each statistic $S$ which we want to investigate, we can calculate
probability distributions of $S$ in the `null' and `bumpy' models.  If $S$
is
a powerful detector of non-gaussianity, there should be little or no
overlap between the two distributions.  Figure 2 shows these distributions
for our favorite statistic, $S_{3;6}$, using
a signal-to-noise ratio of 1.25 (about the same level as seen in
the UCSB SP91 data set).  There is indeed
very little overlap:
the value of $S_{3;6}$ calculated on a `bumpy' data set typically
exceeds the values of all but a small fraction of the null sets.
This
``small fraction'' varies from one `bumpy' set to another, but its
average value is 1.2\% (from now on, we'll refer to this as
``a mean significance of 1.2\%''.)

The performance degrades if instead of using sliding windows, we simply
calculate $S_3$ on the entire data set at once; the mean significance
rises
from 1.2\% to 3.4\%.
For comparison, figure 3 shows the distributions of absolute values of
skewness (calculated
in sliding windows of width 6) of data sets from the null and `bumpy'
models.  The overlap is tremendous; at this noise level, skewness could
never reliably distinguish the two models.  Performance is even worse if
we
calculate skewness on the whole data set instead of on a sliding subset.
Evidently, $S_3$ is a much more powerful detector of non-gaussian bumps
than
is skewness.

To test our assertion that each statistic
$S_q$ is most sensitive to bumps of width of about $q$, we performed
monte carlo runs like those described above for a
variety of `bumpy' models with bumps of different widths.  For each model,
we measured the average significance obtained using $S_{q;3(q-1)}$ for
$q =1,..5.$
 The results are shown in figure 4, which plots mean significance
vs. full-width-half-max of the bump model for each of the five
statistics.  As expected, each statistic $S_q$ reaches its maximum power
(lowest mean significance) for bumps of full-width-half-max near $q$
(or slightly higher.)

\bigskip
\centerline {\bf 4. Optimal Statistics}
\smallskip

We have justified the statistic $S_{q;3(q-1)}$ by an incremental process,
starting with skewness, the simplest detector of non-gaussianity,
and modifying it to counter its obvious shortcomings.  Our monte
carlo results showed that the resulting statistic is a much
better detector of non-gaussian `bumps' than is skewness, but we would
like to go furthur and show that it is optimal or near-optimal
for this job, at least over certain classes of related statistics.

The procedure of calculating $S_{q;3(q-1)}$ can be separated into three steps.
First we convolve the data set with a square tooth of width $q$
(a function equal to one at $q$ adjacent points, and zero elsewhere).
Next we take the third
power of the convolved data points.
Finally we add the results for each connected subset or `window' of
length $3(q-1)$ within the data set, and take the most extreme value of
$S_q$ as our final statistic $S_{q;3(q-1)}$.  For each of these three
steps, we can investigate whether a modification of the procedure would
produce stronger results.

\medskip
\centerline{\bf 4.1  Optimal Choice of Convolution Function}
\smallskip

Our choice to convolve the data with a square tooth
function amounts to a filtered
deconvolution about the gradient signature we are searching for (in this
case a bump),
 with the high-frequency components suppressed.
It is {\it not} better to use the rigorous
deconvolution function of the signature we are seeking; this method is
notoriously vulnerable to high frequency noise. But it {\it is}
worthwhile to see whether convolving the data with some other function,
rather than the arbitrarily chosen square tooth, would produce
a better statistic.

As mentioned before, $S_q$ is nearly equivalent to a linear
combination of several three-point functions.  Except for its
treatment of points near the window edges, $S_3$ is proportional to
\eqn\eeethreeptapprox{\eqalign{
	C_{000} + 2(C_{001} + C_{011}) + 2C_{012} + (C_{002} + C_{022})
}}

To the extent that this approximation holds, the search for the optimal
convolution function
is equivalent to a search for the optimal linear combination of
three-point functions.
We define the generalized three-point function $S_{G3}$ by:
\eqn\eeesg{\eqalign{
	S_{G3}\equiv C_{000}+t(C_{001}+C_{011})+u(C_{012})+v(C_{002}+C_{022})
}}
$C_{001}$ and $C_{011}$
share the same coefficient for reasons of symmetry, as do $C_{002}$ and
$C_{022}$.  There is no coefficient before $C_{000}$ because an overall
multiplicative constant does not affect a statistic's ability to
distinguish between distributions of different {\it shapes.}

$S_{G3}$ includes all six of the three-point functions which involve no
more
than three adjacent points at a time.  Wider-ranging three-point
functions, such as $C_{013}$, are not included because we are attempting
to
generalize $S_3$, which searches most powerfully for bumps spanning 2 or 3
points.

To estimate the optimal coefficients $t$, $u$, and $v$, the procedure is
as follows.
We first adopt two simple analytic models of null (gaussian) and `bumpy'
(non-gaussian) distributions. We then
calculate the mean of $S_{G3}$ for the bumpy model, and its mean and
variance for the null model.
We define the `bump-resolving power' $R$ as the distance, measured in
standard deviations of the null model, between the means of the null and
bumpy models:
\eqn\eeer{\eqalign{
R \equiv {<S_{G3}>_{{bump}} - <S_{G3}>_{{null}} \over
\sqrt {<S_{G3}^2>_{{null}} - <S_{G3}>^2_{{null}}}}
}}

Finally we maximize R as a function of $t$, $u$, and $v$.
The quantity $R$ is not the most accurate measure one could think of,
but is at least straightforwardly calculable.
\bigskip
To simplify the calculation, we work in an infinitely long window of
length $N\rightarrow\infty$, instead of the window of length $L=6$ which
we
shall use in
practice.  This underscores the fact that choosing a statistic (like
$S_q$)
and choosing a window size are two separate ideas; the idea of sliding
windows is not specific to $S_q$ but improves the performance of almost
any
statistic.

\vfill\eject

\medskip
{\noindent \bf 4.1.1 Null Model:}
\smallskip

The null model consists of $N$ points drawn from a gaussian distribution
of zero mean and unit variance, with one important modification: each data
set is  zero mean. If
$\{y_i\}$ are a set of independent points drawn from a gaussian
distribution,
$$
x_i = y_i - {1 \over N} \sum _{j = 1} ^N y_j
$$
This sounds like a trivial change,
especially for large data sets which tend to have means very close to
zero anyway, but the explicit normalization makes a surprisingly large
difference in the calculation even as $N\rightarrow\infty$.  We do
{\it not} explicitly normalize each set to unit variance, because it can
be shown to make no difference in the $N\rightarrow\infty$ limit,.

In order to calculate $R$ from equation \eeer, we need to know $<S_{G3}>$
and $<S_{G3}^2>$ for this
model.  $<S_{G3}>$ is clearly zero, and using other symmetry properties,
$$
<S_{G3}^2> = <C_{000}^2> + 2t^2<C_{001}^2> + u^2<C_{012}^2> +
2v^2<C_{002}^2>
$$

So we need to calculate expectations such as
$$
<C_{000}^2>={1 \over N^2} \sum _{i=1}^N \sum _{j=1}^N
\langle x_i^3x_j^3\rangle
$$
for a gaussian distribution with unit variance.

Expectations such as $<x_i^3x_j^3>$ can be calculated using Wick's
theorem,
starting with
the fundamental 2-point expectation $<x_ix_j> = \delta _{ij} - N^{-1}$.
The $N^{-1}$ term is
due to the normalization of the data $\{x_i\}$ to zero mean.
 The results are:
$$
<C_{000}^2> = 6 N^{-1} \qquad
<C_{001}^2> = 2 N^{-1} \qquad
<C_{002}^2> = 2 N^{-1} \qquad
<C_{012}^2> = N^{-1}
$$

All other terms in the expression for $<S_{G3}^2>$ are zero,
 so
\eqn\eeesgsqnull{\eqalign{
<S_{G3}^2>_{null} \rightarrow (6 + 4t^2 + u^2 + 4v^2) N^{-1}
\qquad \qquad \hbox{as } N\rightarrow\infty
}}
\vfill\eject

\medskip
{\noindent \bf 4.1.2 Bumpy Model:}
\smallskip

To represent non-gaussian, `bumpy' data sets we take
$$
x_n = ay_n + bm_n
$$
where the `noise'  $y_n$ is drawn from a zero mean, unit variance normal
distribution, and $m_n$ is the underlying `bumpy' model:
$$
m_n = g\left( e^{-\alpha (n-n_0)^2}-c \right) \quad \hbox{where }
c = {1 \over N} \sqrt {\pi \over \alpha} \quad \quad
g = N^{1 \over 2} \left( {2 \alpha \over \pi} \right)^{1 \over 4}
$$

The bump center $n_0$ is chosen randomly and is not necessarily an
integer.   The constants $g$ and $c$ are chosen so that, as
$N\rightarrow\infty$, $m_n$
will also have zero mean and unit variance (when averaged over all
$n_0$).  The
purpose of $a$ and $b$ is to set the signal-to-noise ratio: $SNR =
{b\over a}$, and $a^2+ b^2 = 1$.  It is
straightforward to check that all terms involving the noise
term give zero in $S_{G3}$ (this would not be true for higher moments).
Averaging over $n_0$ converts all sums to integrals, which in the limit
$N\rightarrow\infty$
yield the results:
$$
<C_{000}> = b^3N^{1 \over 2} \left( {8 \alpha \over 9\pi} \right)^{1
\over 4} \qquad
<C_{001}> = <C_{000}> e^{-{2 \over 3} \alpha}
$$
$$
<C_{002}> = <C_{000}> e^{-{8 \over 3} \alpha}  \qquad
<C_{001}> = <C_{000}> e^{-2 \alpha}
$$

For $\alpha = 0.5$ (corresponding to a bump with
full-width-half-max of
about 2.8 data points, about the scale we hope to detect with $S_{G3}$),
\eqn\eeesgbump{\eqalign{
<S_{G3}>_{bump} = b^3N^{1 \over 2}[0.6133 + 0.8789t + 0.2256u+0.3233v]
}}

Now we use \eeer\ to estimate the
statistic's power to distinguish gaussian from non-gaussian models:
$$
R \equiv {<S_{G3}>_{{bump}}\over \sqrt {<S_{G3}^2>_{{null}}}} =
b^3N{0.6133 + 0.8789t + 0.2256u+0.3233v \over
\sqrt {6 + 4t^2 + u^2 + 4v^2}}
$$

The optimal values of $t$, $u$, and $v$ are those which maximize $R$;
a numerical search for these yields the optimized three-point statistic:
\eqn\eeesgopt{\eqalign{
S_{G3} = C_{000} + 2.15 (C_{001} + C_{011}) + 2.21 C_{012} + 0.79
(C_{002} + C_{022})
}}

As we hoped, this result is similar to \eeethreeptapprox, confirming that
our original combination of three-point functions (or equivalently, our
choice to convolve the data set with a square tooth) is probably
among the most powerful methods.  Since this calculation was approximate,
 we checked it with
more precise monte carlo runs, which confirmed that no other no other choice
of convolution function gives dramatically better results.  We have settled on
the choice of a square tooth as the best combination of simplicity and
power.

\bigskip
\centerline{\bf 4.2 Optimal choice of power}
\smallskip

After convolving the data set with a square tooth  of width $q$, $S_q$
requires us to sum the third powers of the convolved data points.  We
should investigate whether taking some power other than the third would
give better results.
The higher the power, the more emphasis is given to the most
extreme points in the data set (after convolution.)  Emphasizing extreme
points has the advantage of reducing the effects of noise,  since it
prevents several small bumps, caused by noise, from matching the
effect of a large bump in the signal.  The drawback is that
for a real signal, the highest
points will have neighbors which are also higher than
average, since neither a physical gradient region on the sky nor the
instrument's response function have perfectly sharp edges.  High points
due to extreme values of noise will not in
general have unusual neighbors.
So focusing too heavily on extreme points throws away information
which would help distinguish a physical signal from noise.  The optimal
choice of power is that which balances these two competing effects.

The answer is not obvious and is clearly model-dependent, so we turn
again to monte carlo results.  Consider a class of statistics
\eqn\eeesqp{\eqalign{
S^{(p)}_{3;6} \equiv {1 \over (N-2)} \sum_{i=1}^{N-2}
\left({x_i + x_{i+1} + x_{i+2} \over 3} \right) ^p
}}
These
statistics are calculated in sliding windows of width 6; they differ from
$S_{3;6}$ only in the use of the $p$-th power rather than the third.
We investigated their ability to distinguish two different `bumpy' models
from white noise.  One model, described in the earlier section
on monte carlo results, used bumps of gaussian profile with
full-width-half-max of 2.8 data points.  The other model used bumps of a
square tooth profile with three adjacent, equally high points randomly
placed in the data set.  Gaussian white noise was added to both models at a
signal-to-noise ratio of 1.25.  For the gaussian-profile model, we expect
the resolving strength to peak at some finite power $p$, while the
square bumps should be best resolved at very high $p$ since the argument
for lower powers applies only to bumps in which points near the bump
have
non-zero expectations.

The results are shown in Figure 5, which plots mean significance of
detection vs. $p$.  For detection of the gaussian-profile
bumps, the optimal power was $p=3$.  For the square bumps, higher
powers are always better, as expected.  Again, neither gradient
regions on the CMBR nor instrument response functions are expected to
have sharp cutoffs, so we view the gaussian profiles as more physically
realistic than the square ones, and continue to use $p=3$.  However, the
results show that $p=3$ is only slightly better than other nearby
choices, so we will not hesitate to use $p=4$ later in the paper when we
generalize the statistic to search for other types of gradient signatures
(because even powers will be more convenient
than odd ones).

 Also, we should note that
$S^{(p)}_{3;6}$ becomes very simple as $p \rightarrow \infty$.  Our
statistic is then equivalent (in its relative ranking of
different data sets, which is the only thing that matters) to simply
convolving the data
with a square tooth of width three, and then choosing the most extreme point.
The `sliding window'  becomes irrelevant in this limit.  Readers who feel
that this extra simplicity is worth sacrificing some power may prefer to
use this straightforward procedure.

\bigskip
\centerline {\bf 4.3 Optimizing Window Size}
\smallskip

The final choice that we have made is to use windows of length $L =
3(q-1)$ (equation \eeel).  There is not much of interest to say about this
choice.
We gave a rough justification earlier, and monte carlo runs confirm
that it is the best or nearly the best length for a wide range of $q$
(when searching for bumps of full-width-half-max near $q$.)

\bigskip
\centerline{\bf 5. Experimental Results}
\smallskip

We focus on a run of the UCSB SP91 experiment which observed 13
points in four frequency channels (Gaier {\it et. al.} 1992).
  Cosmological fluctuations should be
frequency-independent,
 so we can average the four channels to better
distinguish cosmological fluctuations from instrument noise and, possibly,
from astrophysical and atmospheric effects. The mean of the four channels
is shown in figure 5.

We applied the statistic $S_{3;6}$ to
data from each of the four channels and to their mean.
We compared the results to G. Efstathiou's 1000
simulations of UCSB SP91's view of standard
inflation skies (after adding
gaussian
noise at the estimated experimental level,
 and removing a
best-fit line, as was done to the actual UCSB SP91 data.)  Channels 1 and 2
(the two lowest-frequency channels, covering 25-27.5 GHz and
 27.5-30 GHz respectively) appear highly non-gaussian.  Both achieve 0.1\%
significance (only 1 of the 1000 inflationary skies
gave as large a value of $S_3$.)  The average of
all four channels does nearly as well, at 0.2\% significance.
Channels 3 and 4 were not conclusively non-gaussian.

By contrast, if we had used pure skewness, we would find
the
mean of the four channels non-gaussian at only 5\% significance.  If we
had used
skewness on the whole data set, instead of with sliding windows, we would
conclude the mean of the four channels
to be non-gaussian at only 11\% significance.

We chose $S_3$ (rather than some other $S_q$) as the preferred statistic
for
analyzing data sets because we expect the perceived size of the gradient
regions to be not much larger than the lower limit set by the instrument
response function (two or three pixels.)  For comparison, Figure 6 shows
the significance levels at which all four channels, as well as their
mean, can be
shown non-gaussian by the various $\{S_q\}$, with $q$ ranging from 1 to
13.  $S_{3;6}$ provides the strongest overall results, although $S_{2;3}$
and $S_{4;9}$ both outperform it on individual channels.

These results show that the UCSB SP91 data is strongly non-gaussian, but
non-gaussian data does not  necessarily  imply non-gaussian cosmology.
The data sets contain a visible spike spanning about 2 pixels (clearly
visible in figure 5.)  This could be the signature of a sharp
gradient generated by a non-gaussian cosmological model, but there are
several other possibilities.  These include galactic foreground sources,
extragalactic but noncosmological sources (unlikely; such sources could
not easily match the spatial structure of the data), or a systematic
instrumental effect such as sidelobe pickup.
With UCSB SP91's limited range of frequency (25-35 GHz) there is not
enough spectral information to reliably distinguish astrophysics from
cosmology (i.e., by fitting to the spectra of synchrotron or bremsstrahlung
radiation). However, if a gaussian theory with small correlations
on two-degree scales, like standard inflation, is to
be believed, we must assume that the signal in
points number 7 and 8  is
non-cosmological, and remove those points from the data
set.  The fluctuations of the remaining eleven points may then
be used to impose constraints on the cosmological fluctuations.
We removed a best-fit line from the remaining points,
because the line already removed from the scan must be
assumed invalid if two points were contaminated.  The remaining points
have a quite reasonable chi-squared of $9.9$, quite
reasonable for 9 degrees of freedom.
We then compared the r.m.s.
to those of the standard inflation simulated
data sets (with points 7 and
8 likewise removed, and a new best-fit line subtracted from the remaining
points). After this procedure, only 1 of the 1000 simulated data sets
had an r.m.s. as low as that of the mean of
UCSB SP91's four channels.  We performed the same procedure,  removing
points 6, 7, 8, and 9 (since points near the bump may also be suspect)
and found the UCSB SP91 data was quieter than all but 6 of the 1000
simulations.

If we are to believe the UCSB SP91 data,
the standard inflation theory is caught on the horns of a
dilemma.  If points 7 and 8 of the UCSB SP91 data are of cosmological origin,
the shape of the data is highly non-gaussian and thus inconsistent with
the theory.  If the two points are contaminated, the measured
r.m.s. is too low at high confidence.  Other gaussian models can be tested in a
similar way. Before claiming that we have rejected any
cosmological model, we must
wait to see if these methods demonstrate non-gaussianity in other
experiments.

\bigskip
\centerline{\bf 6. Other Experiments}
\smallskip

$S_{q; 3(q-1)}$ can be applied without modification to
any single-difference experiment.  We recommend $q=3$ unless the
experiment has a very short distance (well under a
degree) between data points, in which case one should try several larger
values of $q$ to search for high-gradient regions typical of
field ordering theories.

For double- and triple-difference experiments, the characteristic `bump'
marking gradient regions will be replaced by more complicated signatures
representing higher derivatives of this sudden gradient (as seen in
Figure 1).  The ideas
developed here, with some modification, should apply to these experiments
as well.  Recall that the procedure of $S_q$ involves convolving the
data set with a  `square tooth' of width $q$, then adding the third
powers of the convolved data points within each sliding window (see the
section on optimal statistics for a discussion of these steps.)  For more
complicated signatures, we need to modify both the convolution function
and the choice of the third power.

The square tooth  was a natural choice for the convolution
function because it approximates the `bump' signature form which we are
looking for.  We will continue to convolve with a function or width q that
approximates the signature form being sought.
Unfortunately, extra complications arise when the gradient--signature
being sought crosses zero (as it does for all but single-difference
experiments.)  The statistics $\{S_q\}$
designed to search for bumps with width of about $q$ data points, were
fairly powerful for a wide range of other widths as well.  But when
searching for a signature which changes sign, a mismatch of widths can
leave the statistic searching for a form which is {\it orthogonal} to that
actually present, thus cancelling the result.  The
more
zero--crossings the signature contains, the more critical it becomes to
use an
accurate width.  This requires knowledge not only of the instrument
response
function but also of the expected width of the gradient region on the
sky.  The search for non-gaussianity becomes uncomfortably theory-specific.

Even if the
width is chosen perfectly, the convolution of a signature function with
an approximation of itself will yield
several adjacent points which are strongly non-zero but {\it alternate in
sign} as rapidly as the signature itself does.  If we added the third (or
any odd) power of these convolved points, they would cancel one another.
 A simple
solution is to raise the convolved data to the fourth power rather
than the third, suffering a slight decrease in resolving power but
gaining robustness.
  For example, in a third-derivative experiment, such as
Dragovan {\it et. al.}'s
Python (Dragovan {\it et. al.} 1993),
a sharp gradient might be best resolved by a statistic of
the form
$$
S \sim \sum_j (x_j - 2x_{j+1} + x_{j+2})^4
$$

The coefficients of $x_j$, $x_{j+1}$, and $x_{j+2}$ are fairly obvious
guesses,
matched to the expected form of the data (Figure 1).  To detect wider
gradient regions or other signature forms, simply use an appropriate
approximation to the shapes
shown in Figure 1; for example, a bump of width four would respond to
$$
S \sim \sum_j (x_j - x_{j+1} - x_{j+2} + x_j)^4
$$
while in a double-difference experiment, gradient--signature
regions
with width of about three data points would respond well to
$$
S \sim \sum_j (x_j - x_{j+2})^4
$$

There is no need to modify the `sliding window' scheme as we look for
more intricate signature forms; windows of
length $3(q-1)$ still work quite well.

We have performed Monte Carlo simulations which confirm that statistics
such as
these are much more powerful than simple skewness or kurtosis at detecting
the
signatures of sharp gradient regions in double- and triple-difference
experiments.  However, there is another option.  Double- and
triple-difference results are often constructed in stages, starting
with
single-difference data and combining adjacent points.  It may be best to
look for regions of sharp gradient in the original, single-difference
data, where they will appear as simple bumps and can
be detected by the comparatively robust statistics $\{S_q\}$.  The
disadvantage to this approach is that single-difference results may be
more vulnerable to systematic errors; we cannot predict in general which
approach will work best for all experiments.
\vfill\eject
\bigskip
\centerline{\bf 7. Conclusions}
\medskip

We have proposed a class of statistics which should be quite powerful in
detecting a wide range of non-gaussian features in one-dimensional data
sets.  Their greatest potential vulnerability is that gaussian data with
significant correlations on the scale of the spacing between data sets
may be hard to distinguish from non-gaussian forms.  This could occur in
cosmological models with unusually strong power spectra at large angular
scales (Crittenden {\it et. al.} 1993),
 in experiments with a short distance between data points, or
in cases where correlations introduced by the instrument's response
function are similar in form to the non-gaussian `signature' being sought.
Any of these factors may make conclusions harder to draw, but they should
not lead to false rejections of a gaussian theory, as long as the
null data sets accurately model the gaussian theory and the
instrument's properties.

Our analysis of the UCSB SP91 experiment indicates that the
CMBR anisotropy is inconsistent with `standard' inflation.
If, as the theory predicts,
the CMBR fluctuations are gaussian, the nongaussianity of the
data must be due to foreground contamination. If
we discard the contaminated points (those responsible for the
non-gaussian shape) we find a level of fluctuations
significantly smaller than the
theory predicts.  One might worry about possible `conspiracies'
here - if a high fluctuation in the CMBR actually contributed
to the nongaussian `bump', we would throw it away when we removed
the contaminated points. Could this not bias us towards low
amplitudes in the remaining points? No, not in
a theory like standard inflation, because the remaining points
are so weakly correlated with the removed points, that they
still provide a fair sample.
Of course, like any other conclusions drawn from
the still very limited data on CMBR anisotropy,
our conclusion requires confirmation
from further experimental results.

It is also important to check whether current non-gaussian theories
are `nongaussian enough'  to account for
a data set like  USCB SP91.  We have calculated our statistic
$S_{3;6}$ on a small number of
sky maps produced by simulations of cosmic texture (Coulson
{\it et. al.} 1993), with the preliminary result that the distribution
of values of $S_{3;6}$ is indeed substantially broader than that
of standard inflation. Further simulation results
will soon allow a more definite conclusion,
along the lines indicated in the introduction (equation \eaneil).

We conclude that either

\noindent{
i) The CMBR fluctuations  are nongaussian, or
gaussian with a larger coherence angle than standard
inflation. In either case the
anisotropy pattern will hold valuable new information about
the mechanisms of structure formation.}

\noindent
ii) The fluctuations are gaussian with a small
coherence angle, and small, too
small for the standard inflationary theory,
but are overlaid
with significant foreground contamination.

\noindent
iii) The UCSB SP91 data set is not a representative sample
of the microwave sky.

\bigskip
\centerline{\bf Acknowledgements}
\medskip

We thank G. Efstathiou for providing standard
inflation-plus-CDM simulated data sets for
the UCSB SP91 experiment, and
M. Dragovan, L. Page, J. Peebles, J. Ruhl, D. Spergel, S. Staggs,
P. Steinhardt, and D. Wilkinson for helpful conversations.
The work of NT was  supported by
NSF contract PHY90-21984,  and
the David and Lucile Packard Foundation.

\bigskip


\centerline{\bf References}
\bigskip

\noindent{
Bennett, D. P. \& Rhie, S. H. 1992, Livermore preprints.
}

\noindent{
Cheng, E. S. {\it et. al.} 1993, NASA/Goddard Space Flight Center preprint.
}

\noindent{
Coulson, D., Pen, U., \& Turok, N. 1993, Princeton preprint PUP-TH-1393.
}

\noindent{
Crittenden, R. {\it et. al.} 1993, U. Pennsylvania preprint.
}

\noindent{
Dragovan, M. {\it et. al.} 1993, report at AAS meeting, Berkeley.
}

\noindent{
Efstathiou, G. 1990 in {\it The Physics of the Early Universe},
eds. A. Heavens, J. Peacock and A. Davies (SUSSP publications).
}

\noindent{
Gaier, T.  {\it et. al.} 1992, Ap. J. Lett. {\bf 398}, L1.
}

\noindent{
Ganga, K. {\it et. al.} 1993, report at AAAS meeting.
}

\noindent{
Gundersen {\it et. al} 1993, CfPA preprint, Berkeley.
}

\noindent{
Kaiser, N. \& Stebbins, A. 1984, Nature {\bf 310}, 391-393.
}

\noindent{
Luo, X.  \& Schramm, D. 1993, Fermilab preprint.
}

\noindent{
Meinhold, P. {\it et. al.} 1993, CfPA preprint, Berkeley.
}

\noindent{
Moessner, R., Perivaropoulos L.,
\& Brandenberger, R.  1993, Brown preprint.
}

\noindent{
Pen, U., Spergel, D., \& Turok, N. 1993, Princeton preprint PUP-TH-1375.
}

\noindent{
Schuster, J. {\it et. al} 1993, Santa Barbara preprint, Ap. J. Lett. {\it to
appear Aug. 1}.
}

\noindent{
Smoot, G. F. {\it et. al} 1992, Ap. J. {\bf 396}, L1.
}

\noindent{
Turok, N. \& Spergel, D. 1990, Phys. Rev. Lett. {\bf 64}, 2736.
}

\bigskip
\centerline{\bf Figure Captions}
\medskip

{\bf Figure 1:} The characteristic `signatures' which would result from a
step function or region of very sharp gradient in the CMBR, for single-
through triple-difference experiments (that is, experiments whose results
approximate the first through the third derivatives of the CMBR
intensity.)  The purpose of this paper is to develop methods of detecting
these signatures against a background of instrument and other noise.

\medskip
{\bf Figure 2:} Probability distributions of the statistic $S_{3;6}$
(that is, $S_3$ as defined in \eeesq, calculated in `sliding windows' of
length six) for the `bumpy' and null models described in the
section on Monte Carlo results.  The two distributions show very little
overlap, so $S_{3;6}$ appears to be powerful at detecting certain types
of non-gaussianity.

\medskip
{\bf Figure 3:} Probability distributions of skewness, calculated in
`sliding windows' of length six, for the same two models used in Fig. 2.
There is considerable overlap; skewness is much less powerful than
$S_{3;6}$ at distinguishing between data sets drawn from these two models.

\medskip
{\bf Figure 4:} The average significance levels achieved by the
statistics $S_{q;3(q-1)}$ for $q=1...5$, when searching for bumps of
various widths (full-width-half-max ranges from 1 to 10.)  We see that to
detect bumps of width $\sim t$ data points, it is best to choose $q$
roughly equal to (or slightly lower than) $t$.

\medskip
{\bf Figure 5:} The average significance levels achieved by the
statistics
$S^{(p)}_{3;6}$ defined in \eeesqp\ as a function of the power $p$,
when used to discriminate gaussian noise from two different
nongaussian models. One model employs bumps of a gaussian profile,
the other bumps of a square tooth profile. We consider the
former more physically reasonable, and consequently adopt the power
$p=3$ in our subsequent analysis.

\medskip
{\bf Figure 6:} A set of results from the UCSB SP91 experiment.  This shows
the temperature offset $\delta T$, averaged over all frequency channels,
for each of 13 points separated by 2.1 degrees.  UCSB SP91 is a
single-difference experiment, so these values actually represent
(roughly) the difference between the CMBR intensities at two different
points.  The `spike' visible at points 7 and 8 thus suggests a
region of sharp gradient in the CMBR, unless the data are contaminated
at these points.

\medskip
{\bf Figure 7:}
The significance levels at which each of UCSB SP91's four channels,
 as well as the mean of these channels, can
be shown non-gaussian by each of the statistics
 $S_{q;3(q-1)}$ for $1 \leq q \leq 13$.
  $S_{3;6}$ shows the best overall performance.

\bye